\documentclass[12pt]{article}

\usepackage{epsfig} 
\usepackage{amsbsy}

\def\beq{\begin{equation}}
\def\eeq{\end{equation}}

\def\beqn{\begin{eqnarray}}
\def\eeqn{\end{eqnarray}}

\def\({\left(}
\def\){\right)}

\def\Dx{\int {\cal D}x }
\def\Q{{\cal Q} }
\def\P{{\cal P} }
\def\M{{\cal M} }

\def\BUB{\Pi^D}

\newcommand{\G}[1]{\Gamma\left({#1}\right)}

\newcommand{\halfD}{\frac{D}{2}}

\def\Id{I^D_4\(\nu_1,\nu_2,\nu_3,\nu_4;\{Q_i^2\}\)}
\def\Jdone{J^D_{4}\(\{\mu_1,\mu_2,\mu_3,\mu_4\},\mu_5,\mu_6,\mu_7;\{Q_i^2\}\)}
\def\Jdtwo{J^D_{4}\(\mu_1,\{\mu_2,\mu_3,\mu_4,\mu_5\},\mu_6,\mu_7;\{Q_i^2\}\)}
\def\Jdthree{J^D_{4}\(\mu_1,\mu_2,\{\mu_3,\mu_4,\mu_5,\mu_6\},\mu_7;\{Q_i^2\}\)}
\def\Jdfour{J^D_{4}\(\mu_1,\mu_2,\mu_3,\{\mu_4,\mu_5,\mu_6,\mu_7\};\{Q_i^2\}\)}

\def\poch{Pochhammer}

\def\f21{{_2F_1}}

\def\dl#1{$$\displaylines{\quad#1}$$}

\def\li#1{\,{\rm Li}_2\left(#1 \right)}
\def\swap{\(s \leftrightarrow t,\ \nu_1 \leftrightarrow \nu_4,\ \nu_2
\leftrightarrow \nu_3\)}

\def\pow{(-1)^\halfD~}
\def\M1{M^2}
\def\al{\alpha}
\def\bt{\beta}
\def\ga{\gamma}
\def\de{\delta}
\def\ap{\alpha'}
\def\bp{\beta'}
\def\gp{\gamma'}
\def\ep{\epsilon}

\def\dl#1{$$\displaylines{\quad#1}$$}

\def\li#1{\,{\rm Li}_2\left(#1 \right)}
\def\lit#1{\,{\rm Li}_3\left(#1 \right)}
\def\Re{\mathop{\rm Re}}

\catcode`\@=11

\@addtoreset{equation}{section}

\def\theequation{\thesection.\arabic{equation}}

\def\@normalsize{\@setsize\normalsize{15pt}\xiipt\@xiipt
\abovedisplayskip 14pt plus3pt minus3pt%
\belowdisplayskip \abovedisplayskip
\abovedisplayshortskip \z@ plus3pt%
\belowdisplayshortskip 7pt plus3.5pt minus0pt}

\def\small{\@setsize\small{13.6pt}\xipt\@xipt
\abovedisplayskip 13pt plus3pt minus3pt%
\belowdisplayskip \abovedisplayskip
\abovedisplayshortskip \z@ plus3pt%
\belowdisplayshortskip 7pt plus3.5pt minus0pt
\def\@listi{\parsep 4.5pt plus 2pt minus 1pt
     \itemsep \parsep
     \topsep 9pt plus 3pt minus 3pt}}

\@twosidetrue

\catcode`@=12

\catcode`\@=11

\def\section{\@startsection{section}{1}{\z@}{3.5ex plus 1ex minus
   .2ex}{2.3ex plus .2ex}{\large\bf}}

\def\thesection{\arabic{section}}
\def\thesubsection{\arabic{section}.\arabic{subsection}}
\def\thesubsubsection{\arabic{section}.\arabic{subsection}.\arabic{subsubsection}}

\def\appendix{\setcounter{section}{0}
 \def\thesection{\Alph{section}}
 \def\theequation{\Alph{section}.\arabic{equation}}
 \def\thesubsection{\Alph{section}.\arabic{subsection}}
\def\thesubsubsection{\Alph{section}.\arabic{subsection}.\arabic{subsubsection}}

 \def\section{\@startsection{section}{1}{\z@}{3.5ex plus 1ex minus
   .2ex}{2.3ex plus .2ex}{\large\bf}}
}

\def \ep{\epsilon}
\def \to   {\mbox{$\rightarrow$}}

\newcommand\hepph[1]{{\tt hep-ph/#1}}
\newcommand\hepth[1]{{\tt hep-th/#1}}

\def\ord#1{{\cal O}\(#1\)}

\newcount\minutes
\newcount\scratch

\def\timestamp{%
\scratch=\time
\divide\scratch by 60
\edef\hours{\the\scratch}
\multiply\scratch by 60
\minutes=\time
\advance\minutes by -\scratch
---$\,$\hours:\null
\ifnum\minutes< 10 0\fi
\the\minutes}

\begin{document}
\begin{titlepage}
\nopagebreak
{\flushright{
        \begin{minipage}{4cm}
         DTP/99/88 \\
        {\tt hep-ph/9907523}\hfill \\
        \end{minipage}        }

}
\vfill
\begin{center}
{\LARGE \bf \sc
 \baselineskip 0.9cm
 Application of the negative-dimension approach 
to massless scalar box integrals

 }
\vskip
1.3cm 
{\large  C.~Anastasiou\footnote{e-mail: {\tt Ch.Anastasiou@durham.ac.uk}},
E.~W.~N.~Glover\footnote{e-mail: {\tt E.W.N.Glover@durham.ac.uk}}  and
C.~Oleari\footnote{e-mail: {\tt Carlo.Oleari@durham.ac.uk}}} 
\vskip .2cm 
{\it Department of Physics, 
University of Durham, 
Durham DH1 3LE, 
England } 
\vskip
1.3cm    
\end{center}

\nopagebreak
\begin{abstract}

We study massless one-loop box integrals by treating the number of space-time
dimensions $D$ as a negative integer.  We consider integrals with up to three
kinematic scales ($s$, $t$ and either zero or one off-shell legs) and with
arbitrary powers of propagators. For box integrals with $q$ kinematic scales
(where $q=2$ or 3) we immediately obtain a representation of the graph in
terms of a finite sum of generalised hypergeometric functions with $q-1$
variables, valid for general $D$.  Because the power each propagator is
raised to is treated as a parameter, these general expressions are useful in
evaluating certain types of two-loop box integrals which are one-loop
insertions to one-loop box graphs. We present general expressions for this
particular class of two-loop graphs with one off-shell leg, and give explicit
representations in terms of polylogarithms in the on-shell case.

\end{abstract}
\vfill
\vfill
\end{titlepage}
\newpage                                                                     

\section{Introduction} 
\label{sec:intro}
 
Box integrals play an important role in the perturbative description of $2
\,\to\, 2$ scattering processes. Classic examples at one-loop include the
scattering of light-by-light~\cite{LbyL} and the scattering of
partons~\cite{ES}.  Recent improvements of experimental measurements demand
even more precise theoretical predictions and there is significant interest
in determining $2 \,\to\, 2$ cross sections at the two-loop order.  To
achieve this goal requires the evaluation of certain master two-loop graphs,
such as the planar double-box graph~\cite{Smirnov,Smirnov2}, or some one-loop
box integrals with bubble insertions on one of the propagators.

In 1987, Halliday and Ricotta~\cite{HR} suggested a method of calculating
loop integrals based on treating the number of space-time dimensions $D$ as a
negative integer. Because loop integrals are analytic in $D$ (and also in the
powers of the propagators), this is a valid procedure and, although the
intermediate steps may be carried out in negative $D$ (and in particular
series expansions can be made), $D$ remains a parameter of the calculation
and can be taken to be positive after integration. The problem of loop
integration is replaced by that of handling infinite series.
This idea was neglected for some time until Suzuki and Schmidt started a more
systematic application of the negative dimension method (NDIM) to a number of
two-loop integrals~\cite{SS2loop}, three-loop integrals~\cite{SS3loop},
one-loop tensor integrals~\cite{SStensor} as well as the one-loop massive box
integral for the scattering of light by light~\cite{SSbox}.  In this last
paper, Suzuki and Schmidt discovered that as well as reproducing the known
hypergeometric-series representations of Ref.~\cite{Dbox}, valid in
particular kinematic regions, hypergeometric solutions valid in other
kinematic domains are simultaneously obtained.  Of course, all of these
solutions are related by analytic continuation.  However, it is easy to
envisage integrals that yield hypergeometric functions where the analytic
continuation formulae are not known a priori.  In these cases, having series
expansions directly available in all kinematic regions may be very useful.

Recently, we have generalised this method to describe massive $n$-point
one-loop graphs with general powers of the propagators and arbitrary
dimension $D$~\cite{AGO}.  For graphs with $m$ mass scales, $q$ external
momentum scales and $n$ legs, we have written down a template series solution
with $(m+q+n)$ summation indices, together with a linear system of $(n+1)$
constraints.  The template solution is completely general, while the
constraints can be read off the specific Feynman graph.  By solving the
system of constraints, we obtain many solutions with $(m+q-1)$ summation
indices, each of which can be identified directly as a hypergeometric
function in the appropriate convergence region.  The full solution in a
particular kinematic region is formed by adding the solutions that converge
in that region.  It turns out that by keeping the parameters general, it is
easier to identify the regions of convergence of the hypergeometric series
and, therefore, which hypergeometric functions to group together.  This has the
additional advantage of allowing a connection with the general
tensor-reduction program based on integration by parts of Refs.~\cite{Tar1,
Tar2} where the tensor integrals are linear combinations of scalar integrals
with either higher dimension or propagators raised to higher powers.  It is
the goal of this paper to consider massless box integrals and to obtain
expressions in terms of hypergeometric functions valid for general powers of
the propagators and arbitrary dimension.

Our paper is arranged as follows.  In Sec.~\ref{sec:oneloop} we show how NDIM
can be applied to construct the template solutions for one-loop box integrals
together with the linear system of constraints that relates the powers of the
propagators in the loop integral to the summation variables.  We give the
expressions for the solutions in different kinematic regions for massless
scalar box integrals with one off-shell leg and for the on-shell case in terms
of hypergeometric functions of one or two variables. In both cases, $D$ is
arbitrary and the propagators are raised to arbitrary powers. As an application
of the general formulae, in Sec.~\ref{sec:twoloop} we consider a particular
class of two-loop box integrals which are one-loop box graphs with bubble
insertions on one of the legs.  We give general formulae for the scalar
integrals with three powers of propagators set to unity and one propagator
(corresponding to the place where the one-loop insertion is made) kept
arbitrary. In this case, identities amongst hypergeometric functions can be
used to simplify the general expressions. We show how to evaluate the
hypergeometric functions in the on-shell case and, by making a series expansion
in $\ep = (4-D)/2$, give explicit expressions in terms of logarithms and
polylogarithms for the relevant two-loop scalar integrals. Finally, our
findings are summarised in Sec.~\ref{sec:conc}.

\section{The general massless one-loop box integral}
\label{sec:oneloop}

\begin{figure}
\begin{center}
~\epsfig{file=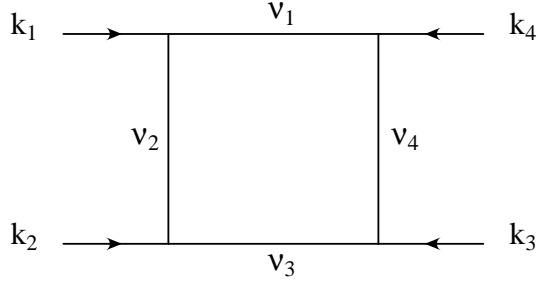,height=4cm}
\end{center}
\caption[]{The one-loop box diagram.}
\label{fig:box}
\end{figure}

The generic massless one-loop box integral in $D$-dimensional Minkowski
space with loop momentum $k$ is given by
\begin{equation} 
\label{eq:IDloop}
\Id = 
\int  \frac{d^Dk}{i\pi^{D/2}} \frac{1}{A_1^{\nu_1}\ldots A_4^{\nu_4}}, 
\end{equation} 
where, as indicated in Fig.~\ref{fig:box},
the external momenta $k_i$ are all incoming so that $\sum_{i=1}^4
k_i^\mu = 0$ and the massless propagators have the form
\begin{eqnarray} 
\label{eq:Ai}
A_1 &=& k^2 + i0, \nonumber \\
A_i &=& \left(k+\sum_{j=1}^{i-1} k_j\right)^2  + i0\qquad {i \neq 1}.
\end{eqnarray}
The external momentum scales are indicated with $\{Q_i^2\}$. In our case they
are the Mandelstam variables $s=\(k_1+k_2\)^2$, $t=\(k_2+k_3\)^2$ and the
external masses $k_i^2= M_i^2$.  In this paper we will focus on box integrals
with at most one off-shell leg, so that we have $k_i^2=0$ for $i=1,2,3$, and
$k_4^2 = \M1$.  For standard integrals, the powers $\nu_i$ to which each
propagator is raised are usually unity. However, we wish to leave the powers
as general as possible. Later on we will use these general expressions to
derive some results for two-loop box integrals with one-loop insertions on
the propagators.

We can rewrite Eq.~(\ref{eq:IDloop}) using Schwinger parameters $x_i$, 
so that
\begin{equation}
\Id = \Dx
\int 
\frac{d^Dk}{i\pi^{D/2}}
~\exp\left (\sum_{i=1}^4 x_iA_i\right),
\label{eq:form1}
\end{equation}
where we have used the shorthand
\begin{equation}
\Dx =(-1)^\sigma\left(\prod_{i=1}^4 \frac{1}{\Gamma(\nu_i)}
\int^{\infty}_0 
dx_i x_i^{\nu_i-1}\right),
\end{equation}
with 
\begin{equation}
\label{eq:sigma}
\sigma = \sum_{i=1}^{4} \nu_i.
\end{equation}
Performing the Gaussian integral in a
straightforward way we have the usual Minkowski-space result for massless
integrals
\begin{equation}
\Id = \Dx
\frac{1}{\P^{D/2}} ~\exp(\Q/\P),
\label{eq:form2}
\end{equation}
where
\begin{equation}
\P = x_1+x_2+x_3+x_4,
\end{equation}
while for box integrals with one off-shell leg ($k_4^2=\M1$)
\begin{equation}
\Q = x_1x_3\, s + x_2 x_4\, t +x_1 x_4\, \M1. 
\end{equation}
As usual, in the physical region $t < 0$ and $s > 0$.

To evaluate the integral further, we adopt the suggestion of Halliday and
Ricotta~\cite{HR} and treat the number of dimensions $D$ as a negative
integer.  This is valid because the loop integral is an analytic function of
$D$. We follow the approach suggested by Suzuki and
Schmidt~\cite{SS2loop}--\cite{SSbox} and detailed in~\cite{AGO} by viewing
Eqs.~(\ref{eq:form1}) and~(\ref{eq:form2}) as existing in negative
dimensions.  We make a series expansion in $x_i$ in both
Eqs.~(\ref{eq:form1}) and~(\ref{eq:form2}).  The role of having $D < 0$ is
that the power of $\P$ is now positive allowing a multinomial expansion.
Following the notation of~\cite{AGO}, we have
\begin{eqnarray}
\lefteqn{I_4^{D}(\nu_1,\nu_2,\nu_3,\nu_4;s,t,\M1) }\nonumber \\
&=& 
\Dx
\sum_{n_1,\ldots,n_4=0}^{\infty}
\int 
\frac{d^Dk}{i\pi^{D/2}}
\frac{(x_1A_1)^{n_1}}{n_1!}
\frac{(x_2A_2)^{n_2}}{n_2!}
\frac{(x_3A_3)^{n_3}}{n_3!}
\frac{(x_4A_4)^{n_4}}{n_4!}\nonumber \\
&=& 
\label{eq:expand}
\Dx
\sum_{{p_1,\ldots,p_4 =0 \atop {q_1,\ldots,q_3 =0 }}}^{\infty}
\frac{\(x_1x_3 s\)^{q_1}\(x_2x_4 t\)^{q_2}\(x_1x_4 \M1\)^{q_3}}{q_1! q_2 ! q_3 !}
\frac{x_1^{p_1}\ldots
x_4^{p_4}}{ p_1!\ldots p_4!}
\, (p_1+p_2+p_3+p_4)!,\phantom{aaa}  
\end{eqnarray}
with the constraint 
\beq 
q_1+q_2+q_3+p_1+p_2+p_3+p_4=-\halfD,
\eeq
that ensures that the power of $\Q$ and $\P$ match up correctly.  The
integers $p_i$ and $q_i$ are introduced in making the multinomial expansions
of $\P$ and $\Q$ respectively.  If more than one leg is off shell, then there
will be additional terms in $\Q$ leading to more summation variables.
Similarly, if we take the $\M1 \,\to\, 0$ limit, this is the same as
fixing $q_3 = 0$ in Eq.~(\ref{eq:expand}).

The $x_i$ are independent variables so that for the
equality~(\ref{eq:expand}) to hold, the integrands themselves must be equal.
Therefore, by selecting the coefficient of the powers of $x_i^{-\nu_i}$,
where $\nu_i = -n_i$, on both sides of the equality we find
\begin{eqnarray}
\label{eq:template}
\lefteqn{I_4^{D}(\nu_1,\nu_2,\nu_3,\nu_4;s,t,\M1) }\nonumber \\
&=& 
\sum_{{p_1,\ldots,p_4 =0 \atop {q_1,\ldots,q_3 =0 }}}^{\infty}
\frac{\G{1+p_1+p_2+p_3+p_4}}{\G{1+q_1}\G{1+q_2}\G{1+q_3}}
\(\prod_{i=1}^4  \frac{\G{1-\nu_i}}{\G{1+p_i}}\)
\,  s^{q_1} t^{q_2}\(\M1\)^{q_3}, 
\end{eqnarray}
subject to the system of constraints
\begin{eqnarray}
\label{eq:sysbox1}
q_1+q_3+p_1 &=& -\nu_1,\nonumber \\
q_2+p_2 &=& -\nu_2,\nonumber \\
q_1+p_3 &=& -\nu_3, \\
q_2+q_3+p_4 &=& -\nu_4,\nonumber \\
q_1+q_2+q_3+p_1+p_2+p_3+p_4 &=& -D/2.\nonumber
\end{eqnarray}
There are seven summation variables and five constraints so that two
variables will be unconstrained.  The procedure for developing the solution
for the loop integral further is detailed in Ref.~\cite{AGO}.  Each of the
fifteen solutions of the system is inserted into the template
solution~(\ref{eq:template}).  For example, solving with respect to the
indices $\{q_1, q_2\}$, we find
\begin{eqnarray*}
p_1 &=& \nu_2+\nu_3+\nu_4+q_2-D/2, \\
p_2 &=& -\nu_2-q_2, \\
p_3 &=& -\nu_3-q_1, \\
p_4 &=& \nu_1+\nu_2+\nu_3+q_1-D/2, \\
q_3 &=& -q_1-q_2+D/2-\nu_1-\nu_2-\nu_3-\nu_4,
\end{eqnarray*}
which is then applied to~(\ref{eq:template}).  $\Gamma$ functions that depend
on the unconstrained variables $q_1$ and $q_2$ are converted into \poch\
symbols
\begin{equation}
\label{eq:poch}
(z,n) \equiv \frac{\Gamma(z+n)}{\Gamma(z)},
\end{equation}
because they are the most suitable way to write generalized hypergeometric
functions.
Denoting this solution as $I^{\{q_1,q_2\}}$ and introducing the shorthand
notation
\beq
\nu_{ij} = \nu_i+\nu_j, \qquad \nu_{ijk} = \nu_i+\nu_j+\nu_k,
\eeq
we have
\begin{eqnarray}
\label{eq:sol1}
I^{\{q_1,q_2\}} &=&  \(\M1\)^{\halfD-\sigma}
\frac{\G{1-\nu_1}\G{1-\nu_4}\G{1+\sigma-D}}
{\G{1+\halfD-\sigma}\G{1+\nu_{123}-\halfD}\G{1+\nu_{234}-\halfD} }\nonumber \\
&\times &
\sum_{q_1,q_2=0}^{\infty}
\frac{\(\sigma-\halfD,q_1+q_2\)\(\nu_3,q_1\)\(\nu_2,q_2\)}
{\(1+\nu_{123}-\halfD,q_1\)\(1+\nu_{234}-\halfD,q_2\) } 
\frac{\(s/\M1\)^{q_1}}{q_1!}
\frac{\(t/\M1\)^{q_2}}{q_2!}.\phantom{aaaaa}
\end{eqnarray}
The second line can be immediately identified as Appell's $F_2$ function (see
Eq.~(\ref{eq:f2_def})) while the apparently divergent $\Gamma$-function
prefactor can be rewritten using the identity
\begin{equation}
\prod_{i=1}^{3}\frac{\Gamma(\alpha_i)}
     {\Gamma(\beta_i)}  = (-1)^{\sum_{i=1}^3 (\beta_i- \alpha_i)}  
\prod_{i=1}^{3}\frac{\Gamma(1-\beta_i)}
{\Gamma(1-\alpha_i)},
\label{eq:flip} 
\end{equation} 
where the index $i$ runs over all of the $\Gamma$ functions in the numerator
and denominator.
This identity holds provided we treat $D/2$ as an integer, as we
have already done in making the multinomial expansion. We see that
\begin{equation} \sum_{i=1}^{3}
(\beta_i- \alpha_i) = \frac{D}{2}, 
\label{eq:args}
\end{equation} 
which is generally true for all solutions and is independent of the $\nu_i$.
Applying~(\ref{eq:flip}) to~(\ref{eq:sol1}) we find that
\begin{eqnarray}
\label{eq:sol}
I^{\{q_1,q_2\}} &=& \pow
\(\M1\)^{\halfD-\sigma}
\frac
{\G{\sigma-\halfD}\G{\halfD-\nu_{123}}\G{\halfD-\nu_{234}} }
{\G{\nu_1}\G{\nu_4}\G{D-\sigma}}
\nonumber \\
&\times &
F_2\(
\sigma-\halfD,\nu_3,\nu_2,1+\nu_{123}-\halfD,1+\nu_{234}-\halfD,
\frac{s}{\M1},\frac{t}{\M1} \).
\end{eqnarray}

Similarly, the other fourteen solutions are given by:
\begin{eqnarray}
I_4^{\{p_1,p_4\}}&=& \pow
s^{\halfD-\nu_{123}}t^{\halfD-\nu_{234}}\(\M1\)^{\nu_{23}-\halfD}\nonumber \\
&& \times
~~~\frac{\G{\nu_{123}-\halfD}\G{\nu_{234}-\halfD}
      \G{\halfD-\nu_{12}}\G{\halfD-\nu_{34}}\G{\halfD-\nu_{23}}}
{\G{\nu_1}\G{\nu_2}\G{\nu_3}\G{\nu_4}\G{D-\sigma}} \nonumber \\
&& \times
~~~{F_2}\left(
\halfD-\nu_{23},\halfD-\nu_{12},\halfD-\nu_{34},1+\halfD-\nu_{123},1+\halfD-\nu_{234},
\frac{s}{\M1},\frac{t}{\M1}\right), \nonumber \\
I_4^{\{p_1,q_1\}}&=& \pow
t^{\halfD-\nu_{234}}\(\M1\)^{-\nu_1}
\frac{\G{\nu_{234}-\halfD}
      \G{\halfD-\nu_{123}}\G{\halfD-\nu_{34}}}
{\G{\nu_2}\G{\nu_4}\G{D-\sigma}} \nonumber \\
&& \times
~~~{F_2}\left(
\nu_1,\nu_3,\halfD-\nu_{34},1+\nu_{123}-\halfD,1+\halfD-\nu_{234},
\frac{s}{\M1},\frac{t}{\M1}\right), \nonumber \\
I_4^{\{p_4,q_2\}}&=& \pow
s^{\halfD-\nu_{123}}\(\M1\)^{-\nu_4}
\frac{\G{\nu_{123}-\halfD} 
      \G{\halfD-\nu_{12}}\G{\halfD-\nu_{234}}}
{\G{\nu_1}\G{\nu_3}\G{D-\sigma}} \nonumber \\
&& \times
~~~{F_2}\left(
\nu_4,\halfD-\nu_{12},\nu_2,1+\halfD-\nu_{123},1+\nu_{234}-\halfD,
\frac{s}{\M1},\frac{t}{\M1}\right), \nonumber \\
I_4^{\{p_2,p_4\}}&=& \pow
s^{\halfD-\nu_{123}}t^{-\nu_2}\(\M1\)^{\nu_2-\nu_4}
\frac{\G{\nu_{123}-\halfD}\G{\nu_4-\nu_2}
      \G{\halfD-\nu_{12}}\G{\halfD-\nu_{34}}}
{\G{\nu_1}\G{\nu_3}\G{\nu_4}\G{D-\sigma}} \nonumber \\
&& \times
~~~{H_2}\left(
\nu_4-\nu_2,\halfD-\nu_{12},\nu_2,\halfD-\nu_{34},1+\halfD-\nu_{123},
\frac{s}{\M1},-\frac{\M1}{t}\right), \nonumber \\
I_4^{\{p_2,q_1\}}&=& \pow
t^{-\nu_2}\(\M1\)^{\halfD-\nu_{134}}
\frac{\G{\nu_{134}-\halfD}
      \G{\halfD-\nu_{123}}\G{\halfD-\nu_{34}}}
{\G{\nu_1}\G{\nu_4}\G{D-\sigma}} \nonumber \\
&& \times
~~~{H_2}\left(
\nu_{134}-\halfD,\nu_3,\nu_2,\halfD-\nu_{34},1+\nu_{123}-\halfD,
\frac{s}{\M1},-\frac{\M1}{t}\right), \nonumber \\
I_4^{\{p_4,q_3\}}&=& \pow
s^{\halfD-\nu_{123}}t^{-\nu_4}
\frac{\G{\nu_{123}-\halfD}\G{\nu_2-\nu_4}
      \G{\halfD-\nu_{12}}\G{\halfD-\nu_{23}}}
{\G{\nu_1}\G{\nu_2}\G{\nu_3}\G{D-\sigma}} \nonumber \\
&& \times
~~~{S_1}\left(
\nu_4,\halfD-\nu_{23},\halfD-\nu_{12},1-\nu_2+\nu_4,1+\halfD-\nu_{123},
-\frac{s}{t},\frac{\M1}{t}\right), \nonumber \\
I_4^{\{q_1,q_3\}}&=& \pow
t^{\halfD-\sigma} 
\frac{\G{\sigma-\halfD}
      \G{\halfD-\nu_{123}}\G{\halfD-\nu_{134}}}
{\G{\nu_2}\G{\nu_4}\G{D-\sigma}} \nonumber \\
&& \times
~~~{S_1}\left(
\sigma-\halfD,\nu_1,\nu_3,1+\nu_{134}-\halfD,1+\nu_{123}-\halfD,
-\frac{s}{t},\frac{\M1}{t}\right),\nonumber \\
I_4^{\{p_2,p_3\}}&=& \pow
s^{-\nu_3}t^{-\nu_2}\(\M1\)^{\halfD-\nu_{14}}
\frac{\G{\nu_{14}-\halfD} 
      \G{\halfD-\nu_{12}}\G{\halfD-\nu_{34}}}
{\G{\nu_1}\G{\nu_4}\G{D-\sigma}} \nonumber \\
&& \times
~~~{F_3}\left(
\nu_2,\nu_3,\halfD-\nu_{34},\halfD-\nu_{12},1+\halfD-\nu_{14},
\frac{\M1}{t},\frac{\M1}{s}\right), \nonumber \\
I_4^{\{p_2,q_3\}}&=& \pow
s^{\halfD-\nu_{134}}t^{-\nu_2}
\frac{\G{\nu_{134}-\halfD}\G{\nu_4-\nu_2}
      \G{\halfD-\nu_{34}}\G{\halfD-\nu_{14}}}
{\G{\nu_1}\G{\nu_3}\G{\nu_4}\G{D-\sigma}} \nonumber \\
&& \times
~~~{S_2}\left(
\nu_{134}-\halfD,\nu_4-\nu_2,\halfD-\nu_{34},\nu_2,1+\nu_{14}-\halfD,
\frac{\M1}{s},\frac{s}{t}\right), \nonumber \\
I_4^{\{p_1,p_3\}}&=&I_4^{\{p_2,p_4\}}\swap ,\nonumber \\
I_4^{\{p_3,q_2\}}&=&I_4^{\{p_2,q_1\}}\swap ,\nonumber \\
I_4^{\{p_1,q_3\}}&=&I_4^{\{p_4,q_3\}}\swap ,\nonumber \\
I_4^{\{q_2,q_3\}}&=&I_4^{\{q_1,q_3\}}\swap ,\nonumber \\
I_4^{\{p_3,q_3\}}&=&I_4^{\{p_2,q_3\}}\swap .
\end{eqnarray}
The definitions of the functions $F_3$, $H_2$, $S_1$ and $S_2$ are given in
Sec.~\ref{subsec:series} together with a table of their regions of
convergence.

We divide the kinematic regions up as shown in Fig.~\ref{fig:region}:
\begin{figure}
\begin{center}
~\epsfig{file=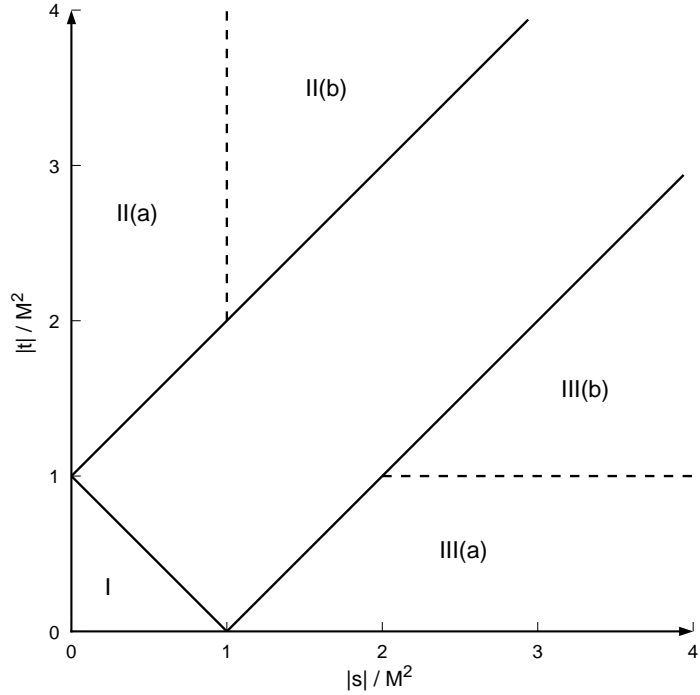,height=10cm}
\end{center}
\caption[]{The kinematic regions for the one-loop box with
one off-shell leg.  The solid line shows the phase-space boundary
$|s|+|t|=\M1$, together with the reflections $|s|=|t|+\M1$ and $|t|=|s|+\M1$.
The reflections are relevant for the convergence properties of the
hypergeometric functions which only involve the absolute values of ratios of
the scales.  The dashed lines show the boundaries $|s|=\M1$ and $|t|=\M1$.}
\label{fig:region}
\end{figure}
\begin{equation}
\label{eq:regions}
\begin{array}{ll}
{\rm region\ I:} & \displaystyle{\M1 > |s|+|t|, }
\nonumber\vspace{1mm} \\  
{\rm region\ II(a):} &  \displaystyle{|t| > \M1+|s| {\rm\ and\ } \M1 > |s|,} 
\nonumber\vspace{1mm}  \\  
{\rm region\ II(b):} &  \displaystyle{|t| > \M1+|s| {\rm\ and\ } |s| > \M1,}
\vspace{1mm}  \\
{\rm region\ III(a):} & \displaystyle{ |s| > \M1+|t| {\rm\ and\ } \M1 > |t|,} 
\nonumber\vspace{1mm}  \\  
{\rm region\ III(b):} & \displaystyle{ |s| > \M1+|t| {\rm\ and\ } |t| > \M1,} 
\nonumber 
\end{array}
\end{equation}
and, applying the convergence criteria of Table~\ref{tab:convergence}
to each of the fifteen solutions, we find
that they are distributed as follows:
\begin{eqnarray}
{\lefteqn{\rm in~region~I}\hspace{6cm}} 
\nonumber \\
\label{eq:sol_box_I}
I_4^D(\nu_1,\nu_2,\nu_3,\nu_4;s,t,\M1)&=&
I_4^{\{ q_1,q_2 \}}+I_4^{\{ p_1,p_4 \}}+
I_4^{\{ p_4,q_2 \}}+I_4^{\{ p_1,q_1 \}},
\\
{\lefteqn{\rm in~region~II(a)}\hspace{6cm}} 
\nonumber \\
\label{eq:sol_box_IIa}
I_4^D(\nu_1,\nu_2,\nu_3,\nu_4;s,t,\M1)&=&
I_4^{\{ p_2,p_4 \}}+I_4^{\{ p_2,q_1 \}}
+I_4^{\{ p_4,q_3 \}}+I_4^{\{ q_1,q_3 \}},
\\
{\lefteqn{\rm in~region~II(b)}\hspace{6cm}} 
\nonumber \\
\label{eq:sol_box_IIb}
I_4^D(\nu_1,\nu_2,\nu_3,\nu_4;s,t,\M1)&=&
I_4^{\{ p_2,p_3 \}}+I_4^{\{ p_2,q_3 \}}
+I_4^{\{ p_4,q_3 \}}+I_4^{\{ q_1,q_3 \}},
\\
{\lefteqn{\rm in~region~III(a)}\hspace{6cm}} 
\nonumber \\
\label{eq:sol_box_IIIa}
I_4^D(\nu_1,\nu_2,\nu_3,\nu_4;s,t,\M1)&=&
I_4^{\{ p_1,p_3 \}}+I_4^{\{ p_3,q_2 \}}
+I_4^{\{ p_1,q_3 \}}+I_4^{\{ q_2,q_3 \}},
\\
{\lefteqn{\rm in~region~III(b)}\hspace{6cm}} 
\nonumber \\
\label{eq:sol_box_IIIb}
I_4^D(\nu_1,\nu_2,\nu_3,\nu_4;s,t,\M1)&=&
I_4^{\{ p_2,p_3 \}}+I_4^{\{ p_3,q_3 \}}
+I_4^{\{ p_1,q_3 \}}+I_4^{\{ q_2,q_3 \}}.\phantom{aaaaaaa}
\end{eqnarray}
Some solutions are convergent in more than one region.  For example,
$I_4^{\{p_4,q_3\}}$ and $I_4^{\{q_1,q_3\}}$ are convergent in both
regions II(a) and II(b) while $I_4^{\{p_2,p_3\}}$ is convergent in both II(b)
and III(b).
We also see that in region II(a), two of the solutions
$\(I_4^{\{p_2,p_4\}}\right.$ and 
$\left. I_4^{\{p_4,q_3\}}\)$ contain dangerous $\Gamma$ functions when $\nu_2 =
\nu_4$.  These divergences indicate the region of a logarithmic
analytic continuation and can be regulated by letting $\nu_2 = \nu_4 +
\delta$, canceling the divergence, and then setting $\delta \,\to\, 0$.
Similarly, the two divergent contributions in region II(b)
$\(I_4^{\{p_2,q_3\}}\right. $ and $\left. I_4^{\{p_4,q_3\}}\right)$ also
cancel in this limit.

We can perform several checks of these results.
\begin{itemize}
\item[-] {\bf Analytic continuation}\\
The solutions in the different regions are related by analytic
continuations of the hypergeometric functions (see for example the appendix
of Ref.~\cite{AGO}). 

\item[-] {\bf The} $\boldsymbol{\nu_i = 0}$ {\bf limit}\\
By pinching out one or more of the propagators (which corresponds to setting
$\nu_i= 0$) we obtain results for triangle or bubble integrals (see
Ref.~\cite{AGO}).  For example, if we set $\nu_2 = \nu_3 = 0 $, then any term
containing $1/\Gamma(\nu_2)$ or $1/\Gamma(\nu_3)$ is eliminated.  In fact,
only five solutions survive, one in each group.  In each case, the
hypergeometric function collapses to unity and we obtain the expected result
for the massless-bubble integral with off-shellness $\M1$ in each of the five
kinematic regions thereby spanning the whole of phase space
\begin{equation}
\label{eq:sol_bub}
I_2^D\(\nu_1,\nu_4;\M1\) = 
\(\M1\)^{\halfD-\nu_{1}-\nu_{4}} \BUB(\nu_1,\nu_4),
\end{equation}
where we have defined, for future reference, 
\beq
\label{eq:bubble}
\BUB(\mu,\mu') = \pow
\frac{\G{\mu+\mu'-\halfD} 
      \G{\halfD-\mu}\G{\halfD-\mu'}}
{\G{\mu}\G{\mu'}\G{D-\mu-\mu'}}.
\eeq

\item[-] {\bf The massless box:}
$\boldsymbol{I_4^D\(\nu_1,\nu_2,\nu_3,\nu_4;s,t,0\) }$ 

The limit $\M1 \,\to\, 0$ can be taken whenever the kinematic region allows
it, that is to say, in regions II(b) and III(b), where $\M1 < |s|$, $\M1< |t|$.
These two regions are related by the symmetry $\swap$, so we focus only on
region II(b).  Only three of the solutions survive, and we have:
\begin{eqnarray}
\label{eq:sol_box0}
\lefteqn{{\rm if\ } |s| < |t|
 \hspace{1.5cm}} \nonumber\\
\lefteqn{I_4^D\(\nu_1,\nu_2,\nu_3,\nu_4;s,t,0\) 
= \left. I_4^{\{ q_1,q_3 \}}\right|_{\M1=0} 
+ \left. I_4^{\{ p_2,q_3 \}}\right|_{\M1=0}
+ \left. I_4^{\{ p_4,q_3 \}}\right|_{\M1=0} }\nonumber \\
&=& \pow 
t^{\frac{D}{2}-\sigma}
\frac{\Gamma\left(\sigma-\frac{D}{2}\right)
\Gamma\left(\frac{D}{2}-\nu_{134}\right)
\Gamma\left(\frac{D}{2}-\nu_{123}\right)}
{\Gamma\left(\nu_2\right)\Gamma\left(\nu_4\right)
\Gamma\left(D-\sigma\right)}\nonumber \\
&&\times
~~~{_3F_2}\left(\nu_1,\nu_3,\sigma-\frac{D}{2},
1+\nu_{134}-\frac{D}{2},1+\nu_{123}-\frac{D}{2},-\frac{s}{t}\right)
\nonumber \\
&+& \pow
s^{\frac{D}{2}-\nu_{123}}t^{-\nu_4}
\frac{\Gamma\left(\nu_{123}-\frac{D}{2}\right)
\Gamma\left(\nu_2-\nu_4\right)
\Gamma\left(\frac{D}{2}-\nu_{23}\right)
\Gamma\left(\frac{D}{2}-\nu_{12}\right)}
{\Gamma\left(\nu_1\right)\Gamma\left(\nu_2\right)
\Gamma\left(\nu_3\right)\Gamma\left(D-\sigma\right)}\nonumber \\
&&\times
~~~{_3F_2}\left(\nu_4,\frac{D}{2}-\nu_{12},\frac{D}{2}-\nu_{23},
1+\nu_4-\nu_2,1+\frac{D}{2}-\nu_{123},
-\frac{s}{t}\right) \nonumber\\
&+& \pow
s^{\frac{D}{2}-\nu_{134}}t^{-\nu_2}
\frac{\Gamma\left(\nu_{134}-\frac{D}{2}\right)
\Gamma\left(\nu_4-\nu_2\right)
\Gamma\left(\frac{D}{2}-\nu_{14}\right)
\Gamma\left(\frac{D}{2}-\nu_{34}\right)}
{\Gamma\left(\nu_1\right)\Gamma\left(\nu_3\right)
\Gamma\left(\nu_4\right)\Gamma\left(D-\sigma\right)}\nonumber \\
&&\times
~~~{_3F_2}\left(\nu_2,\frac{D}{2}-\nu_{14},\frac{D}{2}-\nu_{34},
1-\nu_4+\nu_2,1+\frac{D}{2}-\nu_{134},
-\frac{s}{t}\right).
\end{eqnarray}
Similarly, taking the same $\M1 \,\to\, 0$ limit for
solution~(\ref{eq:sol_box_IIIb}) in region III(b), we find the result valid
when $ |s| > |t|$, which is also obtained by applying the exchanges $(s
\leftrightarrow t$, $\nu_1
\leftrightarrow \nu_4$, $\nu_2 \leftrightarrow \nu_3)$ to
Eq.~(\ref{eq:sol_box0}). Note that we could have obtained the same result by
returning to the template solution~(\ref{eq:template}) with the system of
constraints~(\ref{eq:sysbox1}) and, after setting $q_3 = 0$, solved the
on-shell box directly.  In this case, there are two external scales, $s$ and
$t$, so that there will be six summation variables ($p_1,\ldots,p_4$ and
$q_1$, $q_2$) and five constraints yielding six solutions, three of which
converge when $|s| < |t|$, again yielding Eq.~(\ref{eq:sol_box0}).

As before, there are apparent divergences in the $\Gamma$ functions when $\nu_2
= \nu_4$ that must be regulated.  This is straightforwardly achieved 
for particular values of the parameters by setting $\nu_2 = \nu_4
+\delta$ and making a Taylor expansion.

\item[-] {\bf The} $\boldsymbol{\nu_i=1}$ {\bf limit:} 
$\boldsymbol{I_4^D(1,1,1,1;s,t,\M1)}$

If we set the propagator power equal to one, then all the
groups~(\ref{eq:sol_box_I})--(\ref{eq:sol_box_IIIb}) give the correct answer
\beqn
&&I_4^D(1,1,1,1;s,t,\M1)= \frac{2}{\ep^2}
\frac{\Gamma^2\(1-\ep\)\G{1+\ep}}{\G{1-2\ep}}\frac{1}{st} \Biggl[ 
(-t)^{-\ep} \f21\(1,-\ep,1-\ep,-\frac{u}{s}\) \hspace{0.8cm}\nonumber\\
&& \hspace{2cm} + (-s)^{-\ep} \f21\(1,-\ep,1-\ep,-\frac{u}{t}\)
-(-\M1)^{-\ep} \f21\(-\ep,1,1-\ep,-\frac{\M1 u}{s t}\)
\Biggr],\nonumber\\
\eeqn
where $u$ is defined by $s+t+u = \M1$ and $\ep=(4-D)/2$.  To obtain this
result we have returned to the series representation of the hypergeometric
function and manipulated the series by repeatedly summing with respect to one
summation index to obtain an ${_2F_1}$ function, applied identities to change
the arguments of the ${_2F_1}$ and rewritten the ${_2F_1}$ 
as a series.  Then we sum with respect to the other index, and repeat if
necessary.  Eventually all of the hypergeometric functions of two variables
can be reduced to ${_2F_1}$ functions.
\end{itemize}

\section{Application to two-loop box graphs}
\label{sec:twoloop}

\begin{figure}
\begin{center}
\epsfig{file=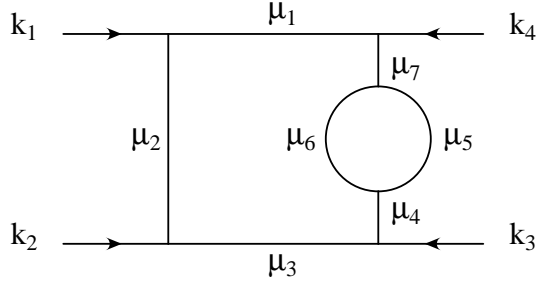,height=4cm}
\end{center}
\caption[]{A one-loop insertion into a one-loop box diagram.}
\label{fig:box2}
\end{figure}

The general results for one-loop box graphs presented in the previous section
may be applied to give analytic results for two-loop box integrals when there
are one-loop insertions on one of the propagators.   As is well known, the
effect of such insertions is to modify the power to which that propagator is
raised. For example, we consider the two-loop integral shown in
Fig.~\ref{fig:box2}, with off-shell legs 
\begin{equation}
\label{eq:twoloop_def}
\Jdfour= 
\int  \frac{d^Dk}{i\pi^{D/2}}
\int  \frac{d^Dl}{i\pi^{D/2}} 
\frac{1}{
A_1^{\mu_1}
A_2^{\mu_2}
A_3^{\mu_3}
A_4^{\mu_4}
B_1^{\mu_5}
B_2^{\mu_6}
A_4^{\mu_7}},
\end{equation}
where the $A_i$ are independent of the second loop momentum $l$ and
are given by Eq.~(\ref{eq:Ai}) while
\begin{eqnarray}
\label{eq:Bi}
B_1 &=& l^2 + i0\nonumber \\
B_2 &=& (l+k+k_1+k_2+k_3)^2 + i0.
\end{eqnarray}
The momentum flowing through the bubble is $k+k_1+k_2+k_3$ so that
the result of the integration over $l$ is (see Eq.~(\ref{eq:sol_bub}))
\begin{equation}
\label{eq:bub}
\int  \frac{d^Dl}{i\pi^{D/2}}\frac{1}{B_1^{\mu_5}B_2^{\mu_6}}
= I_2^D\(\mu_5,\mu_6;A_4\) = \BUB(\mu_5,\mu_6) ~A_4^{\halfD-\mu_{5}-\mu_{6}},
\end{equation}
where $\BUB$ is defined in Eq.~(\ref{eq:bubble}).
In this way, the overall power to which $A_4$ is raised to, in the two-loop
diagram (\ref{eq:twoloop_def}), is
$\mu_4+\mu_5+\mu_6+\mu_7-\halfD$.
Inserting Eq.~(\ref{eq:bub}) into~(\ref{eq:twoloop_def}) we find
\beq
\label{eq:twoloop_res_J}
\Jdfour=
\BUB(\mu_5,\mu_6)~
I_4^D\(\mu_1,\mu_2,\mu_3,\mu_{4567}-\halfD; \{Q_i^2\}\),
\eeq
where $\mu_{4567}=\mu_4+\mu_5+\mu_6+\mu_7$.  Results for diagrams obtained by
pinching out one of the propagators are obtained by setting the corresponding
$\mu_i \,\to\, 0$.  For example, one of the boundary integrals of
Ref.~\cite{Smirnov2} is obtained as the special case
of~(\ref{eq:twoloop_res_J}), with $\mu_4 = \mu_7 = 0$ (see
Fig.~\ref{fig:box3}~(a)).  Similarly, the two-loop diagrams with one-loop
insertions on the other three propagators are defined in an analogous way so
that
\begin{eqnarray}
\Jdone &=& \BUB(\mu_2,\mu_3)~
I_4^D\(\mu_{1234}-\halfD,\mu_5,\mu_6,\mu_7; \{Q_i^2\}\),\nonumber\\
\Jdtwo &=& \BUB(\mu_3,\mu_4)~
I_4^D\(\mu_1,\mu_{2345}-\halfD,\mu_6,\mu_7; \{Q_i^2\}\),\nonumber\\
\Jdthree &=& \BUB(\mu_4,\mu_5)~
I_4^D\(\mu_1,\mu_2,\mu_{3456}-\halfD,\mu_7; \{Q_i^2\}\),\nonumber\\
\end{eqnarray}
where the notation is obvious.

For box graphs with only one off-shell leg, the symmetry of the diagram reduces
the number of distinct integrals to two:
\beqn
\label{eq:onemass_sym}
J^D_{4}\(\{\mu_1,\mu_2,\mu_3,\mu_4\},\mu_5,\mu_6,\mu_7;s,t,\M1\)
&=&J^D_{4}\(\mu_7,\mu_6,\mu_5,\{\mu_4,\mu_3,\mu_2,\mu_1\};t,s,\M1\),\\
J^D_{4}\(\mu_1,\{\mu_2,\mu_3,\mu_4,\mu_5\},\mu_6,\mu_7;s,t,\M1\)
&=&J^D_{4}\(\mu_7,\mu_6,\{\mu_5,\mu_4,\mu_3,\mu_2\},\mu_1;t,s,\M1\),
\phantom{aaa} 
\eeqn
so that it is sufficient to consider diagrams with insertions on the 
third and fourth propagators.
In the on-shell limit $(\M1 \,\to\, 0)$, there is the further relation
\beq
\label{eq:nomass_sym}
J^D_{4}\(\mu_1,\mu_2,\{\mu_3,\mu_4,\mu_5,\mu_6\},\mu_7;s,t\)
=
J^D_{4}\(\mu_7,\mu_1,\mu_2,\{\mu_3,\mu_4,\mu_5,\mu_6\};t,s\),
\eeq
so that for the massless box we only need to consider insertions on a single
propagator.

\subsection{One-loop insertions in the  one-loop box with one off-shell leg}

In this section, we further specify the values of the propagators in the
general forms for the one-loop box graphs of
Eqs.~(\ref{eq:sol_box_I})--(\ref{eq:sol_box_IIIb}): we fix three of the
propagator powers equal to one, while the fourth power is kept free.  Because
of the symmetry properties of the integral~(\ref{eq:onemass_sym}), we need
only to keep either $\nu_4$ or $\nu_3$ general.

\begin{itemize}
\item[-] $\boldsymbol{I_4^D\(1,1,1,\nu_4; s,t,\M1\)}$\\
This limit is appropriate for two-loop diagrams such as that depicted in
Fig.~\ref{fig:box2}.  We choose to work with the solutions in region I, given
by Eq.~(\ref{eq:sol_box_I}).\footnote{ Although we start from the solution
for $|s|+|t| < \M1$, the same expressions can be obtained starting from
any of the other kinematic regions.}  Each of the four solutions is an Appell
$F_2$ function which can be represented as a double Eulerian integral (see
Eq.~(\ref{eq:f2_integral})). 
However, for this choice of the parameters, the $F_2$ functions simplify (see
Refs.~\cite{AGO,erdelyi,KdF}) and we find
\begin{eqnarray}
\label{eq:onemass_nu4}
\lefteqn{I_4^D\(1,1,1,\nu_4; s,t,\M1\) } \nonumber \\
&=& \pow
\(\M1\)^{\halfD-2-\nu_4}\(\M1-t\)^{-1}
\frac
{\G{3+\nu_4-\halfD}\G{\halfD-3}\G{\halfD-2-\nu_4} }
{\G{\nu_4}\G{D-3-\nu_4}}
\nonumber \\
&&\times 
~~~F_1\(1,2+\nu_4-\halfD,1,4-\halfD,
\frac{s}{\M1},\frac{s}{\M1-t} \)\nonumber \\
&+& \pow
s^{\halfD-3}t^{\halfD-2-\nu_4}\(\M1-s-t\)^{2-\halfD}\nonumber \\
&& \times
~~~\frac{\G{3-\halfD}\G{2+\nu_4-\halfD}
      \G{\halfD-2}^2\G{\halfD-1-\nu_4}}
{\G{\nu_4}\G{D-3-\nu_4}} \nonumber \\
&+& \pow
t^{\halfD-2-\nu_4}\(\M1-t\)^{-1}
\frac{\G{2+\nu_4-\halfD}
      \G{\halfD-3}\G{\halfD-1-\nu_4}}
{\G{\nu_4}\G{D-3-\nu_4}} \nonumber \\
&& \times
~~~{_2F_1}\left(1,1,4-\halfD,\frac{s}{\M1-t}\right) \nonumber \\
&+& \pow
s^{\halfD-3}\(\M1-s\)^{-\nu_4}
\frac{\G{3-\halfD} 
      \G{\halfD-2}\G{\halfD-2-\nu_4}}
{\G{D-3-\nu_4}} \nonumber \\
&& \times
~~~{_2F_1}\left(1,\nu_4,3+\nu_4-\halfD,\frac{t}{\M1-s}\right). 
\end{eqnarray}
Note that the value of $D$ plays no role in simplifying the hypergeometric
functions and the result given here is for general $D$.  The remaining
hypergeometric functions can now be manipulated using standard identities and
the one-dimensional integral representations given in
Sec.~\ref{subsec:integral} can be used for specific evaluations.  At this
stage, a series expansion in $\ep = (4-D)/2$ becomes necessary.

\item[-]  $\boldsymbol{I_4^D\(1,1,\nu_3,1; s,t,\M1\)}$\\
Similarly, for the case where $\nu_1=\nu_2=\nu_4=1$ and $\nu_3$ is kept
general, we obtain
\begin{eqnarray}
\label{eq:onemass_nu3}
\lefteqn{I_4^D\(1,1,\nu_3,1; s,t,\M1\) }\nonumber \\ 
&=& \pow
\(\M1\)^{\halfD-2} \(\M1-s\)^{-\nu_3} \(\M1-t\)^{-1}
\frac
{\G{3+\nu_3-\halfD}\G{\halfD-2-\nu_{3}}^2 }
{\G{D-3-\nu_3}}
\nonumber \\
&& \times
~~~{_2F_1}\(1,\nu_3,3+\nu_{3}-\halfD,\frac{st}{(\M1-s)(\M1-t)} \)\nonumber \\
&+& \pow
s^{\halfD-2-\nu_{3}}t^{\halfD-2-\nu_3}\(\M1-s-t\)^{2-\halfD}
\(\M1-t\)^{\nu_3-1}
\nonumber \\
&& \times
~~~\frac{\G{2+\nu_{3}-\halfD}^2
      \G{\halfD-2}\G{\halfD-1-\nu_3}^2}
{\G{\nu_3}\G{D-3-\nu_3}} \nonumber \\
&+&  \pow
t^{\halfD-2-\nu_3}\(\M1-t\)^{-1}
\frac{\G{2+\nu_3-\halfD}
      \G{\halfD-2-\nu_{3}}\G{\halfD-1-\nu_3}}
{\G{D-3-\nu_3}} \nonumber \\
&& \times
~~~{_2F_1}\left(1,\nu_3,3+\nu_{3}-\halfD,\frac{s}{\M1-t}\right) \nonumber \\
&+&  \pow
s^{\halfD-2-\nu_{3}}\(\M1\)^{-1}
\frac{\G{2+\nu_{3}-\halfD} 
      \G{\halfD-2}\G{\halfD-2-\nu_3}}
{\G{\nu_3}\G{D-3-\nu_3}} \nonumber \\
&& \times
~~~{F_2}\left(
1,\halfD-2,1,\halfD-1-\nu_{3},3+\nu_3-\halfD,
\frac{s}{\M1},\frac{t}{\M1}\right).
\end{eqnarray}
In this case, one $F_2$ function does not reduce simply and it is necessary
to resort to the two-dimensional integral representation of
Eq.~(\ref{eq:f2_integral}) for explicit evaluation.

\end{itemize}

\subsection{One-loop insertions in the massless one-loop box}

\begin{figure}
\begin{center}
(a)
~\epsfig{file=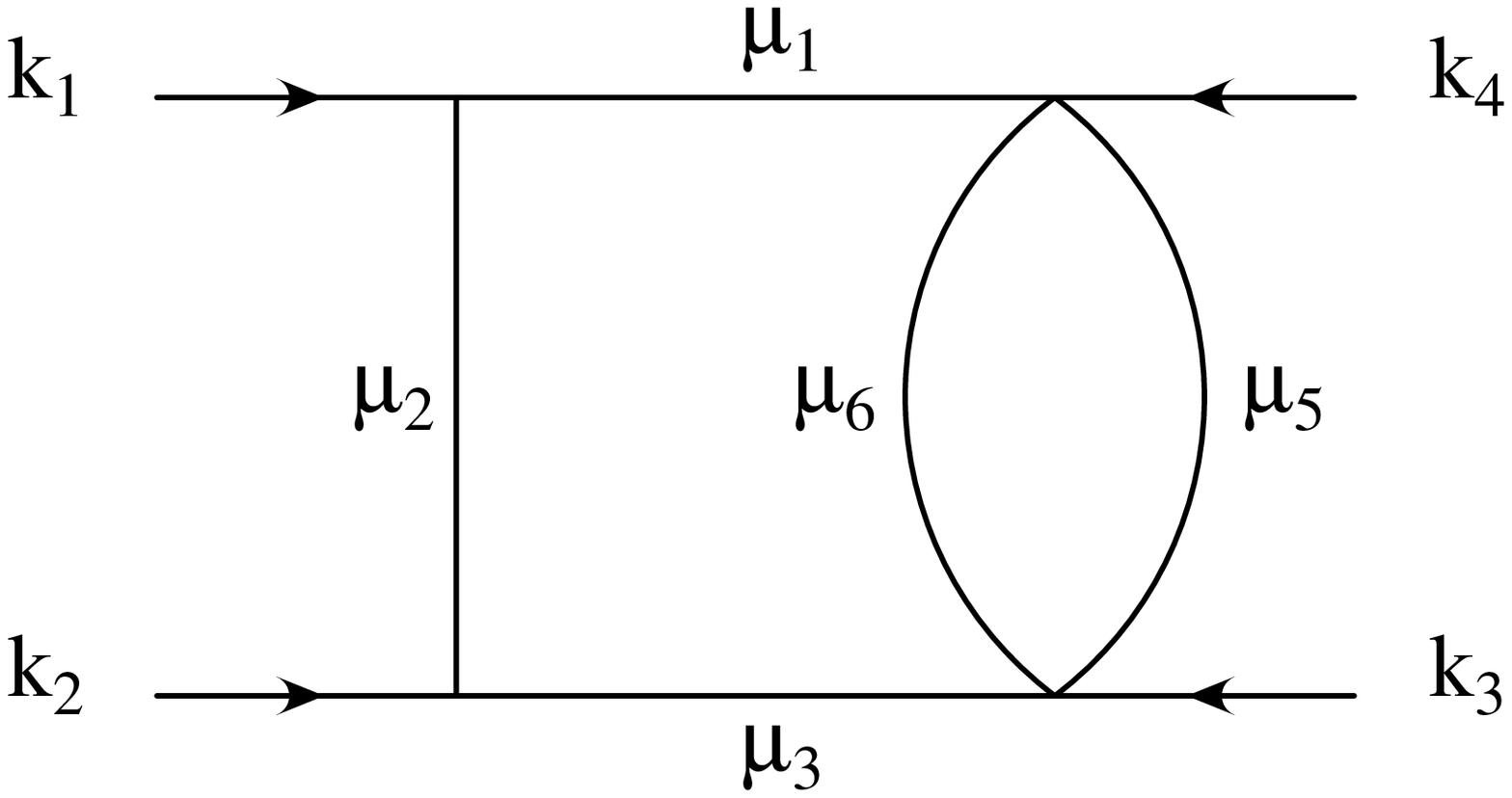,height=4cm}
(b)
~\epsfig{file=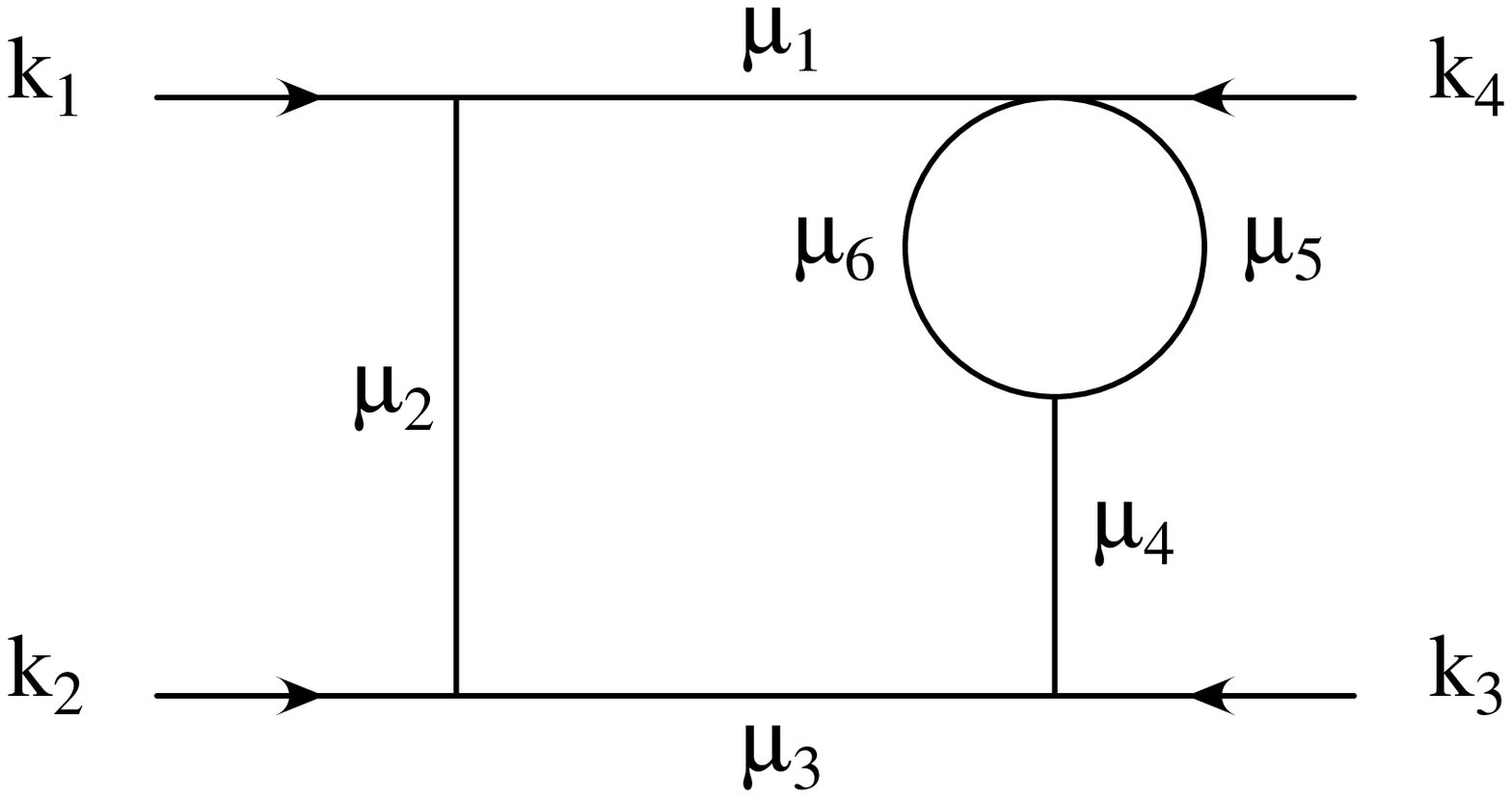,height=4cm}
\end{center}
\caption[]{Two-loop box diagrams with pinched propagators.}
\label{fig:box3}
\end{figure}

We can also attack the problem in the on-shell box.  Here we set
$\nu_1=\nu_2=\nu_3=1$ and keep $\nu_4$ general, which is appropriate for
diagrams such as those shown in Figs.~\ref{fig:box2}
and~\ref{fig:box3}. Insertions on the other legs are given by the symmetry
properties of the integral (see
Eqs.~(\ref{eq:onemass_sym})--(\ref{eq:nomass_sym})).  We therefore choose to
work with the solution valid when $|s| < |t|$ since that contains no $\Gamma$
functions that are singular when $\nu_1 = \nu_3$.  In every case, the
${_3F_2}$ functions of Eq.~(\ref{eq:sol_box0}) reduce to ${_2F_1}$ functions
and we find
\begin{eqnarray}
I_4^D\(1,1,1,\nu_4; s,t\)
&=& \pow 
t^{\frac{D}{2}-3-\nu_4}
\frac{\Gamma\left(3+\nu_4-\frac{D}{2}\right)
\Gamma\left(\frac{D}{2}-2-\nu_4\right)
\Gamma\left(\frac{D}{2}-3\right)}
{\Gamma\left(\nu_4\right)
\Gamma\left(D-3-\nu_4\right)}\nonumber \\
&&\times
~~~{_2F_1}\left(1,1,4-\frac{D}{2},-\frac{s}{t}\right)
\nonumber \\
&+& \pow
s^{\frac{D}{2}-2-\nu_4}t^{-1}
\frac{\Gamma\left(2+\nu_4-\frac{D}{2}\right)
\Gamma\left(\nu_4-1\right)
\Gamma\left(\frac{D}{2}-1-\nu_4\right)^2}
{
\Gamma\left(\nu_4\right)\Gamma\left(D-3-\nu_4\right)}\nonumber \\
&&\times
~~~{_2F_1}\left(1,\frac{D}{2}-1-\nu_4,2-\nu_4,-\frac{s}{t}\right)\nonumber \\
&+& \pow
s^{\frac{D}{2}-3}t^{-\nu_4}
\frac{\Gamma\left(3-\frac{D}{2}\right)
\Gamma\left(1-\nu_4\right)
\Gamma\left(\frac{D}{2}-2\right)^2}
{\Gamma\left(D-3-\nu_4\right)}\nonumber \\
&&\times
~~~\(1+\frac{s}{t}\)^{2-\halfD}.
\end{eqnarray}
There is still an apparent divergence as $\nu_4 \,\to\, 1$ which can be easily
removed by manipulating the hypergeometric functions using the well known
analytic continuations to obtain
\begin{eqnarray}
I_4^D\(1,1,1,\nu_4; s,t\)
\!\!&=& \!\!\pow 
t^{\frac{D}{2}-2-\nu_4}s^{-1}
\frac{\Gamma\left(3+\nu_4-\frac{D}{2}\right)
\G{\halfD-2}\G{2-\halfD}
\Gamma\left(\frac{D}{2}-2-\nu_4\right)}
{
\Gamma\left(\nu_4\right)\Gamma\left(D-3-\nu_4\right) \G{3-\halfD}}\nonumber \\
&&\times
~~~{_2F_1}\left(1,\halfD-2,\halfD-1,\frac{s+t}{s}\right) \nonumber \\
&+&\!\! \pow
s^{\frac{D}{2}-3-\nu_4}
\frac{\G{2+\nu_4-\halfD}
\G{\halfD-\nu_4-1}^2
\G{2-\halfD}}
{\G{D-3-\nu_4}\G{3-\halfD}}\nonumber \\
&&\times
~~~{_2F_1}\left(1,\nu_4,\frac{D}{2}-1,\frac{s+t}{s}\right).
\label{eq:nomass_nu4}
\end{eqnarray}
We have checked that the same result can be obtained by starting from the
general solution valid for $|t| < |s|$.  In this case, we must regulate the
singularity as $\nu_3 \,\to\, \nu_1$ by setting $\nu_3 = \nu_1+\delta$.  The
singularity as $\delta \,\to\, 0$ is canceled by analytically continuing the
${_2F_1}$'s and, after taking the $\delta \,\to\, 0$ limit, we recover
Eq.~(\ref{eq:nomass_nu4}).

\subsubsection{Explicit evaluation of two-loop box integrals}
 
To give more explicit expressions requires a more precise knowledge of
$\nu_4$.  For the two-loop diagram shown in Fig.~\ref{fig:box2} the value of
$\nu_4$ is given by $\mu_4+\mu_5+\mu_6+\mu_7-\halfD=n-\halfD$, where $n$ is an
integer.  The case $n=2$ corresponds to the simplest case $\mu_4=\mu_7 = 0$
and $\mu_5=\mu_6=1$, shown in Fig.~\ref{fig:box3}~(a). Substituting this value
in the general expression (\ref{eq:nomass_nu4}) and restoring all overall
factors, we find that the expression for this two-loop integral in $D=4-2\ep$
is given by (see Eq.~(\ref{eq:twoloop_res_J}))
\begin{eqnarray}
\label{eq:nomass_res1}
J_4^D(1,1,1,\{0,1,1,0\};s,t)
&=&
\( -t\)^{-2\ep}
\frac{K_1}{2\,s \,\ep^3}
~{_2F_1}\(1,-\ep,1-\ep,\frac{s+t}{s}\) \nonumber \\
&+&
\( -s\)^{-2\ep}
\frac{K_2}{2\,s\, \ep^3}
~{_2F_1}\(1,\ep,1-\ep,\frac{s+t}{s}\) ,
\end{eqnarray}
where the constants $K_1$ and $K_2$ are given by
\begin{eqnarray}
K_1 &=& 
\frac{\G{1+2\ep}\G{1-\ep}^3}{\(1-2\ep\)\G{1-3\ep}}
\\
K_2 &=&  
\frac{\G{1+2\ep}\G{1-2\ep}\G{1+\ep}\G{1-\ep}^2}{\(1-2\ep\)\G{1-3\ep}}.
\end{eqnarray}
Note that by starting off with the NDIM approach, we have not actually had to
perform any integrations to reach this result or make any assumptions about
the smallness of $\ep$.  The hypergeometric functions have one-dimensional
integral representations (see Eq.~(\ref{eq:f21_integral})) and can be
expanded around $\ep = 0$ in terms of polylogarithms.  The necessary
integrals are easily done
\beqn
\label{eq:f21gen}
&&\hspace{-0.6cm}\f21\(1,-\ep,1-\ep,x\) =  1+\ep \log(1-x) -\ep^2 \li{x} -\ep^3 \lit{x} +
\ord{\ep^4}\\
\label{eq:f21nu4}
&&\hspace{-0.6cm}\f21\(1,\ep,1-\ep,x\)  =  1- \ep \log(1-x) + 
\ep^2\left[-2\li{\frac{x}{x-1}}-\li{x}\right] \nonumber\\
&&\hspace{-0.6cm} \phantom{\f21\(1,\ep,1-\ep,\)\,\,\,\,}
- \ep^3\left[ 2\lit{\frac{x}{x-1}}+ \lit{x}+ \frac{1}{3} \log^3(1-x) 
\right]\! +\ord{\ep^4},
\eeqn
where the polylogarithms are defined by
\begin{equation}
\label{eq:def_li2}
\li{x} = -\int_0^x dz \, \frac{\log(1-z)}{z} \hspace{1cm} x \leq 1\;,
\end{equation}
and
\beq
\label{eq:def_li3}
\lit{x} = \int_0^1 dz \, \frac{\log(z) \log(1-x z)}{z} = \int_0^x dz \,
\frac{\li{z}}{z}  \hspace{1cm} x \leq 1\;.
\eeq
For $x > 1$, the following analytic continuations should be used
\beqn
\label{eq:continuaz_dilog}
 \li{x\pm i0} &=& -\li{\frac{1}{x}} -\frac{1}{2}
 \log^2 x +\frac{\pi^2}{3} \pm i \pi \log x \hspace{1 cm} x > 1\;, \\
\label{eq:continuaz_trilog}
\lit{x\pm i0} &=& \lit{\frac{1}{x}} -\frac{1}{6}\log^3(x) + \frac{\pi^2}{3}
\log(x) \pm i \frac{\pi}{2} \log^2(x)  \hspace{1 cm} x > 1\;.
\eeqn

Similarly, the integral with only one pinched propagator, $\mu_4=\mu_5=\mu_6=1$
and $\mu_7=0$, shown in Fig.~\ref{fig:box3}~(b) is given by
\begin{eqnarray}
\label{eq:nomass_res2}
J_4^D(1,1,1,\{1,1,1,0\};s,t)
&=&
\( -t\)^{-2\ep}
\frac{3\,K_1}{2\,s\,t\, \ep^3}
~{_2F_1}\(1,-\ep,1-\ep,\frac{s+t}{s}\) \nonumber \\
&-&
\( -s\)^{-2\ep}
\frac{3\,K_2}{4\,s^2 \,\ep^3}
~{_2F_1}\(1,1+\ep,1-\ep,\frac{s+t}{s}\).
\end{eqnarray}
The series expansion for the first hypergeometric function is given by
Eq.~(\ref{eq:f21gen}) while the second can be obtained from
Eq.~(\ref{eq:f21nu4}) by using Gauss's relation between contiguous
hypergeometric functions
\beq
(\bt-\al) (1-x) \, \f21(\al,\bt,\ga,x) -(\ga-\al) \, \f21(\al-1,\bt,\ga,x) + 
(\ga-\bt) \, \f21(\al,\bt-1,\ga,x) =0,
\eeq
such that
\beq
\label{eq:contig}
\f21(1,\bt+1,1-\ep,x) = -\frac{\ep}{\bt (1-x)} + \frac{(\ep+\bt)}{\bt (1-x)} \,
\f21(1,\bt,1-\ep,x). 
\eeq

Finally, the scalar integral for the bubble insertion 
$\mu_4=\mu_5=\mu_6=\mu_7=1$ shown in Fig.~\ref{fig:box2} is
\begin{eqnarray}
\label{eq:nomass_res3}
J_4^D(1,1,1,\{1,1,1,1\};s,t) 
&=&
\( -t\)^{-2\ep}
\frac{3\,(1+3\ep)\,K_1}{2\,(1+\ep)\, s\,t^2 \,\ep^3}
~{_2F_1}\(1,-\ep,1-\ep,\frac{s+t}{s}\) \nonumber \\
&+&
\( -s\)^{-2\ep}
\frac{3\,(1+3\ep)\, K_2}{4\,(1+2\ep)\, s^3 \,\ep^3}
~{_2F_1}\(1,2+\ep,1-\ep,\frac{s+t}{s}\). \phantom{aaa}
\end{eqnarray}
Once again, the series expansion for the first hypergeometric function is
given by Eq.~(\ref{eq:f21gen}) while the second can be obtained
from~Eq.~(\ref{eq:f21nu4}) by repeated use of Eq.~(\ref{eq:contig}).

\section{Conclusions} 
\label{sec:conc}

In this paper we have evaluated one-loop massless box integrals with arbitrary
powers of the propagators and with up to one off-shell leg as combinations of
hypergeometric functions.  The method we have used, first suggested by Halliday
and Ricotta, has its roots in the analytic properties of loop integrals and, in
particular, the possibility of treating the space-time dimensions $D$ as a
negative integer in intermediate steps.  In Ref.~\cite{AGO} we have developed a
general strategy for evaluating one-loop integrals in NDIM and we have pointed
out some subtleties that can occur in the application of the method.  For the
box integrals we have considered here, with $q$ energy scales, we have
expressed the final result as finite sums of hypergeometric functions with
$q-1$ variables, that converge in the appropriate kinematic regions.  The
general results for one off-shell leg in the kinematic regions specified by
Eq.~(\ref{eq:regions}) are given in
Eqs.~(\ref{eq:sol_box_I})--(\ref{eq:sol_box_IIIb}).  Similar expressions for
the on-shell case are given in Eq.~(\ref{eq:sol_box0}).  We would like to point
out that no integration was actually necessary in obtaining these results.

All of these expressions are valid for arbitrary powers of the propagators
and are therefore relevant to classes of multiloop graphs where there are
(multiple) one-loop insertions on the propagators.  We have studied how these
expressions are relevant to this type of two-loop graph and, in particular,
two-loop graphs with three powers of propagators set to unity and one
propagator (corresponding to the place where the one-loop insertion is made)
kept arbitrary. With this choice of parameters, identities amongst
hypergeometric functions can be used to simplify the general expressions.
Explicit results in terms of hypergeometric functions are given for the one
off-shell case in Eqs.~(\ref{eq:onemass_nu4}) and (\ref{eq:onemass_nu3}).  In
the on-shell case, the two-loop scalar integrals reduce down to two Gaussian
${_2F_1}$ functions.  Up to this point we have not actually had to perform
any integrations explicitly or make a series expansion in $\ep =
(4-D)/2$. However, to write the hypergeometric integrals in terms of
logarithms and polylogarithms it is necessary to use an integral
representation and make the series expansion in $\ep$.  For the ${_2F_1}$
functions, the integral representation is one-dimensional and the integrals
are well known.  Explicit results for the graphs of Figs.~\ref{fig:box2}
and~\ref{fig:box3} are given in Eqs.~(\ref{eq:nomass_res1}),
(\ref{eq:nomass_res2}) and~(\ref{eq:nomass_res3}).

It is clear that NDIM is an extremely efficient way of solving one-loop
integrals.   Furthermore, as we have shown in this paper and as Suzuki and
Schmidt~\cite{SS2loop,SS3loop}  have previously shown, NDIM can help in
evaluating multi-loop integrals where there are one-loop insertions on one or
more of the propagators. Whether or not NDIM can provide some non-trivial
results for multi-loop graphs is an interesting, but still open, question.

\section*{Acknowledgements}
We thank J.V. Armitage for stimulating discussions and insights into the field
of Hypergeometric functions and P. Watson and M. Zimmer for useful
conversations.   We acknowledge the assistance of D. Broadhurst and A.
Davydychev regarding generalised hypergeometric functions. C.A. acknowledges
the financial support of the Greek government and C.O.  acknowledges the
financial support of the INFN.

\appendix
\section{Hypergeometric definitions and identities} 
\label{sec:app}

In Sec.~\ref{subsec:series} we give the definitions of the hypergeometric
functions as a series together with their regions of convergence. Integral
representations for the ${_2F_1}$, $F_1$ and $F_2$  functions are given in
Sec.~\ref{subsec:integral} while identities for reducing the $F_1$ and $F_2$
functions to simpler functions are given in Sec.~\ref{subsec:identities}.

\subsection{Series representations} 
\label{subsec:series}

The hypergeometric functions of one variable are sums of \poch\ symbols over
a single summation parameter $m$
\begin{eqnarray}
\label{eq:f21_def}
{_2F_1}\(\al, \bt, \ga,x\) 
&=& \sum_{m=0}^{\infty} \frac{(\al,m)(\bt,m)
}{(\ga,m)} \,\frac{x^m}{m!} 
\\
\label{eq:f32_def}
{_3F_2}\(\al, \bt, \bp ,\ga,\gp, x\) 
&=& \sum_{m=0}^{\infty} \frac{(\al,m)(\bt,m)(\bp,m)
}{(\ga,m)(\gp,m)} \,\frac{x^m}{m!}, 
\end{eqnarray}
which are convergent when $|x| < 1$.

The hypergeometric functions of two variables 
can be written as sums over the integers $m$ and $n$:  $F_i$, $i=1,\ldots,4$
are the Appell functions, $H_2$ a Horn function and $S_1$ and $S_2$ generalised
Kamp\'{e} de F\'{e}riet functions:
\begin{eqnarray}
\label{eq:f1_def}
F_1\(\al, \bt, \bp, \ga,x,y\) 
&=& \sum_{m,n=0}^{\infty} \frac{(\al,m+n)(\bt,m)(\bp,n)
}{(\ga,m+n)} \,\frac{x^m}{m!} \,\frac{y^n}{n!}
\\
\label{eq:f2_def}
F_2\(\al,\bt,\bp,\ga,\gp,x,y \) 
&=& \sum_{m,n=0}^{\infty} \frac{(\al,m+n)(\bt,m)(\bp,n)
}{(\ga,m)(\gp,n)} \,\frac{x^m}{m!}\, \frac{y^n}{n!}
\\
\label{eq:f3_def}
F_3\(\al,\ap,\bt,\bp,\ga,x,y \)
&=& \sum_{m,n=0}^{\infty} \frac{(\al,m)(\ap,n)(\bt,m)(\bp,n)
}{(\ga,m+n)} \, \frac{x^m}{m!}\, \frac{y^n}{n!} 
\\
\label{eq:f4_def}
F_4\(\al,\bt,\ga,\gp,x,y\) 
&=& \sum_{m,n=0}^{\infty} \frac{(\al,m+n)(\bt,m+n)}
{(\ga,m)(\gp,n)} \, \frac{x^m}{m!} \, \frac{y^n}{n!} 
\\
\label{eq:h2_def}
H_2\(\al,\bt,\ga,\gp,\de,x,y\) 
&=& \sum_{m,n=0}^{\infty} \frac{(\al,m-n)(\bt,m)(\ga,n)(\gp,n)}
{(\de,m)}\, \frac{x^m}{m!} \, \frac{y^n}{n!} 
\\
\label{eq:s1_def}
S_1\(\al,\ap,\bt,\ga,\de,x,y\)
&=& \sum_{m,n=0}^{\infty} \frac{(\al,m+n)(\ap,m+n)(\bt,m)}
{(\ga,m+n)(\de,m)}\, \frac{x^m}{m!} \, \frac{y^n}{n!} 
\\
\label{eq:s2_def}
S_2\(\al,\ap,\bt,\bp,\ga,x,y\)
&=& \sum_{m,n=0}^{\infty} \frac{(\al,m-n)(\ap,m-n)(\bt,n)(\bp,n)}{(\ga,m-n)}
\, \frac{x^m}{m!} \, \frac{y^n}{n!}. 
 \phantom{aaa}
\end{eqnarray}

These series converge according to the criteria collected in
Table~\ref{tab:convergence}.
\begin{table}[ht]
\begin{center}
\begin{tabular}{c| c}
 \hline
Function  &  Convergence criteria \\
\hline
$F_1$, $F_3$ & $\displaystyle{|x| < 1}, \ |y| < 1$ \\
$F_2$, $S_1$ & $\displaystyle{|x| + |y| < 1}$ \\
$F_4$ & $\displaystyle{\sqrt{|x|} + \sqrt{|y|} < 1}$ \\
$H_2$, $S_2$ & $ -|x| + 
1/|y| 
> 1$, $|x| <1$, $|y|<1$ \vspace{1mm}  \\
\hline
\end{tabular}
\caption{Convergence regions for some hypergeometric functions of two
variables.}
\label{tab:convergence}
\end{center}
\end{table}
The domain of convergence of the Appell and Horn functions are well known.  
That one for $S_1$ and $S_2$ may be worked out using Horns general
theory of convergence~\cite{Exton}.

\subsection{Integral representations} 
\label{subsec:integral}

Euler integral representations of ${_2F_1}$, $F_1$ and $F_2$ 
are well known~\cite{erdelyi}--\cite{bailey} and we list the relevant formulae
here. 
\dl{
{_2F_1}\(\al,\bt,\ga,x\) = \frac{\G{\ga}}{\G{\bt}\G{\ga-\bt}}
\times \int_0^1 du  \, u^{\bt-1}
(1-u)^{\ga-\bt-1} (1-u x)^{-\al}
\hfill}
\beq
\label{eq:f21_integral}
\Re(\bt)>0, \quad \Re(\ga-\bt) >0.
\eeq
\dl{
F_1(\al, \bt, \bp, \ga,x,y) 
= \frac{\G{\ga}}{\G{\al}\G{\ga-\al}} \int_0^1 du \, 
u^{\al-1}(1-u)^{\ga-\al-1}(1-ux)^{-\bt}(1-uy)^{-\bp}
}
\beq
\label{eq:f1_integral}
 \Re(\al) > 0, \quad \Re(\ga-\al) > 0.
\eeq
\dl{
F_2\(\al,\bt,\bp, \ga,\gp,x,y\) = \frac{\G{\ga}\G{\gp}}{\G{\bt}\G{\bp}
\G{\ga-\bt} \G{\gp-\bp}}\hfill\cr
\phantom{F_2\(\al,\bt,\bp, \ga,\gp,x,y\) =}
\times \int_0^1 du \int_0^1 dv \, u^{\bt-1} v^{\bp-1}
(1-u)^{\ga-\bt-1} (1-v)^{\gp-\bp-1} (1-ux-vy)^{-\al}\hfill
}
\beq
\label{eq:f2_integral}
\Re(\bt)>0, \quad \Re(\bp) > 0, \quad \Re(\ga-\bt) > 0, \quad \Re(\gp-\bp)>0.
\eeq

\subsection{Identities amongst the hypergeometric functions} 
\label{subsec:identities}

The $F_1$ and $F_2$ functions have the following reduction formulae which
leave a single remaining Euler integral at most~\cite{erdelyi}--\cite{bailey}:
\beqn
&&F_1\( \al,\bt,\bp,\bt+\bp,x,y \) = 
(1-y)^{-\al}\f21\(\al,\bt,\bt+\bp, \frac{x-y}{1-y} \) \\
&&F_2\(\al,\bt,\bp,\ga,\al,x,y\) = (1-y)^{-\bp}
F_1\(\bt,\al-\bp,\bp,\ga,x,\frac{x}{1-y} \) \\
&&F_2\(\al,\bt,\bp,\al,\gp,x,y\) = (1-x)^{-\bt}
F_1\(\bp,\bt,\al-\bt,\gp,\frac{y}{1-x},y\) \\
&&F_2\( \al,\bt,\bp,\bt,\gp,x,y\) = (1-x)^{-\al} \f21\(\al,\bp,\gp,
\frac{y}{1-x}\) \\
&&F_2\( \al,\bt,\bp,\al,\al,x,y\) = (1-x)^{-\bt} (1-y)^{-\bp}
\f21\(\bt,\bp,\al,\frac{xy}{(1-x)(1-y)}\)\\
&&F_2\( \al,\bt,\bp,\al,\bp,x,y\) = (1-y)^{\bt-\al} (1-x-y)^{-\bt}
\\
&&F_2\( \al,\bt,\bp,\bt,\bp,x,y\) = (1-x-y)^{-\al}.
\eeqn


\relax
\def\pl#1#2#3{{\it Phys.\ Lett.\ }{\bf #1}\ (19#2)\ #3}
\def\zp#1#2#3{{\it Z.\ Phys.\ }{\bf #1}\ (19#2)\ #3}
\def\prl#1#2#3{{\it Phys.\ Rev.\ Lett.\ }{\bf #1}\ (19#2)\ #3}
\def\rmp#1#2#3{{\it Rev.\ Mod.\ Phys.\ }{\bf#1}\ (19#2)\ #3}
\def\prep#1#2#3{{\it Phys.\ Rep.\ }{\bf #1}\ (19#2)\ #3}
\def\pr#1#2#3{{\it Phys.\ Rev.\ }{\bf #1}\ (19#2)\ #3}
\def\np#1#2#3{{\it Nucl.\ Phys.\ }{\bf #1}\ (19#2)\ #3}
\def\sjnp#1#2#3{{\it Sov.\ J.\ Nucl.\ Phys.\ }{\bf #1}\ (19#2)\ #3}
\def\app#1#2#3{{\it Acta Phys.\ Polon.\ }{\bf #1}\ (19#2)\ #3}
\def\jmp#1#2#3{{\it J.\ Math.\ Phys.\ }{\bf #1}\ (19#2)\ #3}
\def\jp#1#2#3{{\it J.\ Phys.\ }{\bf #1}\ (19#2)\ #3}
\def\nc#1#2#3{{\it Nuovo Cim.\ }{\bf #1}\ (19#2)\ #3}
\def\lnc#1#2#3{{\it Lett.\ Nuovo Cim.\ }{\bf #1}\ (19#2)\ #3}
\def\rnc#1#2#3{{\it Riv.\ Nuovo Cim.\ }{\bf #1}\ (19#2)\ #3}
\def\ptp#1#2#3{{\it Progr. Theor. Phys.\ }{\bf #1}\ (19#2)\ #3}
\def\tmf#1#2#3{{\it Teor.\ Mat.\ Fiz.\ }{\bf #1}\ (19#2)\ #3}
\def\tmp#1#2#3{{\it Theor.\ Math.\ Phys.\ }{\bf #1}\ (19#2)\ #3}
\def\jhep#1#2#3{{\it J.\ High\ Energy\ Phys.\ }{\bf #1}\ (19#2)\ #3}
\def\epj#1#2#3{{\it Eur.\ Phys. J.\ }{\bf #1}\ (19#2)\ #3}
\relax

\end{document}